\documentclass{article}
\usepackage{amsmath,amssymb}
\usepackage{authblk}
\usepackage{cite}
\usepackage{comment}
\usepackage{physics}
\usepackage{mathtools}

\makeatletter
\@addtoreset{equation}{section}

\makeatother

\title{\textbf{Supersymmetric and Gauge-Invariant Path Integral Measure in $\mathcal N=2$ SQCD}}
\author{Akihisa D.-E. Tateishi}
\affil{\it RIKEN, 2-1 Hirosawa, Wako, Saitama 351-0198, Japan}
\date{}

\begin{document}
\maketitle
\begin{abstract}
We define $\mathcal N=2$ supersymmetric and gauge-invariant path integral measure in $D=4$, $\mathcal N=2$ SQCD in terms of $\mathcal N=1$ superfields. As a further consequence, we derive the $\mathcal N=2$ version of the chiral anomaly in an $\mathcal{N}=2$ supersymmetry-preserving manner.
\end{abstract}

\newpage

\section{Introduction}
Supersymmetric field theory has often been considered an object of interest. Especially, $\mathcal N=2$ supersymmetric theory enjoys some intriguing properties. For example, $\mathcal N=2$ chiral matter theory has hyper-K\"ahler structure as its target manifold\cite{9,10,11,12,13}, and supersymmetric localisation theory has special property in the case of $\mathcal N=2$\cite{15,16,17,18}. It is convenient to describe supersymmetric theories using superfields, and in this paper, $\mathcal N=2$ supersymmetric theory is described by $\mathcal N=1$ superfields.

On the other hand, explicit calculation of path integral is one significant way to comprehend quantum field theory. For example, the chiral anomaly in gauge theory is calculated by explicitly executing path integral\cite{1}, and so is its $\mathcal N=1$ supersymmetric version\cite{2,3}. In addition, supersymmetric gauge theory is more deeply understood by explicit calculation of path integral, which is called supersymmetric localisation theory\cite{15,16,17,18}.

In this paper, we define $\mathcal N=2$ supersymmetric and gauge-invariant path integral measure in $D=4$, $\mathcal N=2$ supersymmetric gauge theory (supersymmetric quantum chromodynamics, SQCD) in terms of $\mathcal N=1$ superfields. As a further consequence, we derive the $\mathcal N=2$ version of the chiral anomaly in a way that preserves $\mathcal{N}=2$ supersymmetry.

This paper is organised as follows. In Section \ref{sec:sqcd} we review $\mathcal N=2$ SQCD described by $\mathcal N=1$ superfields. In Section \ref{sec:measure} we define $\mathcal N=2$ supersymmetric and gauge-invariant path integral measure in $\mathcal N=2$ SQCD in terms of $\mathcal N=1$ superfields. In Section \ref{sec:anomaly} we derive the $\mathcal N=2$ SUSY version of chiral anomaly in an $\mathcal{N}=2$ supersymmetry-preserving manner with reference to the previous researches\cite{1,2,3,0}, as a further consequence using the result in Section \ref{sec:measure}. Section \ref{sec:summary} is the summary.

\section{$\mathcal N=2$ SQCD: Review}\label{sec:sqcd}
In this section we review the minimal theory of $D=4$, $\mathcal N=2$ SQCD. The action integral\cite{8,9,10,11,12,13} is, in terms of $\mathcal N=1$ superfields\cite{5},
\begin{gather}
S_{\mathrm{SQCD}}=S_{\mathrm{gauge}}+S_{\mathrm{matter}},\\
S_{\mathrm{gauge}}=\tr\left(\frac14\int d^6z\,W^\alpha W_\alpha+\frac14\int d^6\bar z\,\bar W_{\dot\alpha}\bar W^{\dot\alpha}+\int d^8z\,e^{-V}\Phi^\dagger e^V\Phi\right),\label{eq:lagrangian_gauge}\\
S_{\mathrm{matter}}=\int d^8z\,\left(Z^\dagger e^VZ+\tilde Ze^{-V}\tilde Z^\dagger\right)+\int d^6z\,\tilde Z\Phi Z+\int d^6\bar z\,Z^\dagger\Phi^\dagger\tilde Z^\dagger,\label{eq:lagrangian_matter}
\end{gather}
where $Z$ and $\tilde Z$ (resp. $Z^\dagger$ and $\tilde Z^\dagger$) are chiral (antichiral) matter superfields, i.e.,
\begin{gather}
\bar D_{\dot\alpha}Z=\bar D_{\dot\alpha}\tilde Z=D^\alpha Z^\dagger=D^\alpha\tilde Z^\dagger=0,
\end{gather}
$V$ is a real gauge superfield, accompanied by its superpartner, chiral $\Phi$:
\begin{gather}
V=V^\dagger,\qquad\bar D_{\dot\alpha}\Phi=D^\alpha\Phi^\dagger=0,
\end{gather}
and $W_\alpha$ ($\bar W_{\dot\alpha}$) is the chiral (antichiral) gauge field strength of $V$:
\begin{gather}
W^\alpha\equiv-\frac14\bar D^2\left(e^{-V}D^\alpha e^V\right),\qquad\bar W_{\dot\alpha}\equiv+\frac14D^2\left(e^V\bar D_{\dot\alpha}e^{-V}\right).
\end{gather}
In (\ref{eq:lagrangian_gauge}) and (\ref{eq:lagrangian_matter}), the gauge fields $(V,\Phi)$ are in adjoint represantation, while the matter $Z$'s are in fundamental represantation. 
The actions (\ref{eq:lagrangian_gauge}) and (\ref{eq:lagrangian_matter}), respectively, are manifestly $\mathcal N=1$ supersymmetric, and invariant under gauge transformation
\begin{align}
Z&\mapsto e^{i\Lambda}Z,&Z^\dagger&\mapsto Z^\dagger e^{-i\Lambda^\dagger},\\
\tilde Z&\mapsto\tilde Ze^{-i\Lambda},&\tilde Z^\dagger&\mapsto e^{i\Lambda^\dagger}\tilde Z^\dagger,\\
e^V&\mapsto e^{i\Lambda^\dagger}e^Ve^{-i\Lambda},&e^{-V}&\mapsto e^{i\Lambda}e^{-V}e^{-i\Lambda^\dagger},\\
\Phi&\mapsto e^{i\Lambda}\Phi e^{-i\Lambda},&\Phi^\dagger&\mapsto e^{i\Lambda^\dagger}\Phi^\dagger e^{-i\Lambda^\dagger},
\end{align}
with a chiral gauge transformation parameter $\Lambda$. Furthermore (\ref{eq:lagrangian_gauge}) and (\ref{eq:lagrangian_matter}) are invariant under the second SUSY transformation\footnote{
In order to describe the second SUSY, it suffices that $\delta Z=-(1/4)\bar D^2(\bar\epsilon e^{-V}\tilde Z^\dagger)$ etc., but it does not work afterwards.
}\cite{12}
\begin{align}
\delta Z&=-\frac14\bar D^2\left[(\epsilon+\bar\epsilon)e^{-V}\tilde Z^\dagger\right],& \delta Z^\dagger&=-\frac14D^2\left[(\epsilon+\bar\epsilon)\tilde Ze^{-V}\right],\\
\delta\tilde Z&=+\frac14\bar D^2\left[(\epsilon+\bar\epsilon)Z^\dagger e^V\right],&\delta\tilde Z^\dagger&=+\frac14D^2\left[(\epsilon+\bar\epsilon)e^VZ\right],
\end{align}
\begin{gather}
\delta e^V=(\epsilon+\bar\epsilon)\left(\Phi^\dagger e^V+e^V\Phi\right),\qquad \delta e^{-V}=-(\epsilon+\bar\epsilon)\left(e^{-V}\Phi^\dagger+\Phi e^{-V}\right),\\
\delta\Phi=-\frac12W^\alpha D_\alpha\epsilon=\frac18\bar D^2\left(e^{-V}D^\alpha e^VD_\alpha\epsilon\right),\quad\delta\Phi^\dagger=-\frac12\bar W_{\dot\alpha}\bar D^{\dot\alpha}\bar\epsilon,
\end{gather}
with a global SUSY transformation parameter $\epsilon$ and $\bar\epsilon$, i.e.,
\begin{gather}
\bar D_{\dot\alpha}\epsilon=D^2\epsilon=\partial_a\epsilon=D^\alpha\bar\epsilon=\bar D^2\bar\epsilon=\partial_a\bar\epsilon=0.
\end{gather}

\section{Supersymmetric and gauge-invariant path integral measure}\label{sec:measure}
In this section we define $\mathcal N=2$ supersymmetric and gauge-invariant path integral measure in $\mathcal N=2$ SQCD in terms of $\mathcal N=1$ superfields. As for the path integral of chiral matter fields
\begin{align}
\int [dZd\tilde Z^\dagger dZ^\dagger d\tilde Z]\,e^{-S[Z,\tilde Z^\dagger,Z^\dagger,\tilde Z]},
\end{align}
define the path integral measure
\begin{align}
[dZd\tilde Z^\dagger]\coloneqq\prod_nda_n,\quad
\begin{pmatrix}
Z\\\tilde Z^\dagger
\end{pmatrix}
=\sum_na_n
\begin{pmatrix}
Z_n\\\tilde Z^\dagger_n
\end{pmatrix},
\quad P
\begin{pmatrix}
Z_n\\\tilde Z^\dagger_n
\end{pmatrix}=\lambda_n
\begin{pmatrix}
Z_n\\\tilde Z^\dagger_n
\end{pmatrix},\label{eq:measure}
\end{align}
where 
\begin{align}
P&=
\begin{pmatrix}
-\Phi&-\displaystyle\frac14\bar D^2e^{-V}\\[2ex]
-\displaystyle\frac14D^2e^V&-\Phi^\dagger
\end{pmatrix}
\begin{pmatrix}
\Phi&-\displaystyle\frac14\bar D^2e^{-V}\\[2ex]
-\displaystyle\frac14D^2e^V&\Phi^\dagger
\end{pmatrix}
\\&=
\begin{pmatrix}
\displaystyle\frac1{16}\bar D^2e^{-V}D^2e^V-\Phi^2&\displaystyle\Phi\frac14\bar D^2e^{-V}-\frac14\bar D^2e^{-V}\Phi^\dagger\\[2ex]
-\displaystyle\frac14D^2e^V\Phi+\Phi^\dagger\frac14D^2e^V&\displaystyle\frac1{16}D^2e^V\bar D^2e^{-V}-\Phi^{\dagger2}
\end{pmatrix},
\end{align}
and $\lambda_n$ is the eigenvalue of $P$. The conjugate $[dZ^\dagger d\tilde Z]$ is defined similarly. The measure (\ref{eq:measure}) is not only manifestly $\mathcal N=1$ supersymmetric, but also gauge invariant and $\mathcal N=2$ supersymmetric. In fact, under gauge transformation, the fields are covariantly transformed:
\begin{align}
\begin{pmatrix}
Z\\\tilde Z^\dagger
\end{pmatrix}
\mapsto
\begin{pmatrix}
e^{i\Lambda}&0\\
0&e^{i\Lambda^\dagger}
\end{pmatrix}
\begin{pmatrix}
Z\\\tilde Z^\dagger
\end{pmatrix}
,\qquad
P
\begin{pmatrix}
Z\\\tilde Z^\dagger
\end{pmatrix}
\mapsto
\begin{pmatrix}
e^{i\Lambda}&0\\
0&e^{i\Lambda^\dagger}
\end{pmatrix}
P
\begin{pmatrix}
Z\\\tilde Z^\dagger
\end{pmatrix},
\end{align}
and thus the eigenfunctions are also covariant:
\begin{align}
\begin{pmatrix}
Z_n\\\tilde Z^\dagger_n
\end{pmatrix}
\mapsto
\begin{pmatrix}
e^{i\Lambda}&0\\
0&e^{i\Lambda^\dagger}
\end{pmatrix}
\begin{pmatrix}
Z_n\\\tilde Z^\dagger_n
\end{pmatrix}
,\qquad
P
\begin{pmatrix}
Z_n\\\tilde Z^\dagger_n
\end{pmatrix}
\mapsto
\begin{pmatrix}
e^{i\Lambda}&0\\
0&e^{i\Lambda^\dagger}
\end{pmatrix}
P
\begin{pmatrix}
Z_n\\\tilde Z^\dagger_n
\end{pmatrix}.
\end{align}
Therefore the expansion coefficient $a_n$ is invariant under gauge transformation.
With regard to the second SUSY transformation, similarly, the fields are covariantly transformed:
\begin{align}
\delta
\begin{pmatrix}
Z\\[2ex]\tilde Z^\dagger
\end{pmatrix}
=
\begin{pmatrix}
0&-\displaystyle\frac14\bar D^2(\epsilon+\bar\epsilon)e^{-V}\\[2ex]
+\displaystyle\frac14D^2(\epsilon+\bar\epsilon)e^V&0
\end{pmatrix}
\begin{pmatrix}
Z\\[2ex]\tilde Z^\dagger
\end{pmatrix}
\equiv
\Delta
\begin{pmatrix}
Z\\[2ex]\tilde Z^\dagger
\end{pmatrix},
\end{align}
\begin{align}
\delta\left[
P
\begin{pmatrix}
Z\\\tilde Z^\dagger
\end{pmatrix}
\right]=\Delta\left[
P
\begin{pmatrix}
Z\\\tilde Z^\dagger
\end{pmatrix}
\right],
\end{align}
and therefore the coefficient $a_n$ is also invariant under the second SUSY.
\subsection{Orthonormality of the basis}
With respect to an Hermite inner product
\begin{align}
\braket {a}{b}=\int d^8z\,\left(A^\dagger e^VB+\tilde A e^{-V}\tilde B^\dagger\right)+\int d^6z\,\tilde A\Phi B+\int d^6\bar z\,A^\dagger\Phi^\dagger\tilde B^\dagger,
\end{align}
where
\begin{align}
\ket a:=
\begin{pmatrix}
A\\\tilde A^\dagger
\end{pmatrix},
\end{align}
the operator $P$ is Hermite:
\begin{align}
\mel{a}{P}{b}=\mel{b}{P}{a}^\dagger.
\end{align}
Thus the eigenvectors form a complete and orthonormal basis:
\begin{align}
\braket{\Xi_m}{\Xi_n}=\delta_{mn},\qquad\mathrm{where}\quad
\ket{\Xi_n}:=
\begin{pmatrix}
Z_n\\\tilde Z_n^\dagger
\end{pmatrix},
\end{align}
\begin{align}
\sum_n\ket{\Xi_n(z)}\bra{\Xi_n(z')}&=\sum_n
\begin{pmatrix}
Z_n(z)\\\tilde Z^\dagger_n(z)
\end{pmatrix}
\begin{pmatrix}
Z^\dagger_n(z')&\tilde Z_n(z')
\end{pmatrix}
\begin{pmatrix}
e^V&\Phi^\dagger\\
\Phi&e^{-V}
\end{pmatrix}\\
&=
\begin{pmatrix}
\delta^6(z_L-z_L')&0\\
0&\delta^6(z_R-z_R')
\end{pmatrix},
\end{align}
where $z_L\equiv(x+i\theta\sigma\bar\theta,\theta), z_R\equiv(x-i\theta\sigma\bar\theta,\bar\theta)$.

\section{Chiral anomaly in $\mathcal N=2$ SQCD}\label{sec:anomaly}
In this section we derive the $\mathcal N=2$ SUSY version of chiral anomaly in an $\mathcal{N}=2$ supersymmetry-preserving manner in terms of $\mathcal N=1$ superfields with reference to the previous researches\cite{1,2,3,0}, as a result derived from the path integral measure in the previous section. First, consider the path integral
\begin{align}
\int [dZd\tilde Z^\dagger dZ^\dagger d\tilde Z]\,e^{-S[Z,\tilde Z^\dagger,Z^\dagger,\tilde Z]},
\end{align}
where $S$ is $S_{\mathrm{matter}}$ in (\ref{eq:lagrangian_matter}),
\begin{align}
S=\int d^8z\,\left(Z^\dagger e^VZ+\tilde Ze^{-V}\tilde Z^\dagger\right)+\int d^6z\,\tilde Z\Phi Z+\int d^6\bar z\,Z^\dagger\Phi^\dagger\tilde Z^\dagger.
\end{align}
This action is invariant under chiral transformation
\begin{align}
Z&\mapsto e^{i\lambda}Z,&Z^\dagger&\mapsto Z^\dagger e^{-i\lambda},\\
\tilde Z&\mapsto\tilde Ze^{-i\lambda},&\tilde Z^\dagger&\mapsto e^{i\lambda}\tilde Z^\dagger,
\end{align}
where $\lambda$ is a constant. Next, promote this global chiral transformation to the local one
\begin{align}
Z&\mapsto e^{i\Lambda}Z,&Z^\dagger&\mapsto Z^\dagger e^{-i\Lambda^\dagger},\\
\tilde Z&\mapsto\tilde Ze^{-i\Lambda},&\tilde Z^\dagger&\mapsto e^{i\Lambda^\dagger}\tilde Z^\dagger,
\end{align}
with $\bar D_{\dot\alpha}\Lambda=D^2\Lambda=0$, and consider the variation of the path intergral
\begin{align}
&\int [dZd\tilde Z^\dagger dZ^\dagger d\tilde Z]\,e^{-S[Z,\tilde Z^\dagger,Z^\dagger,\tilde Z]}\\
=&
\int [dZ'd\tilde Z'^\dagger dZ'^\dagger d\tilde Z']\,e^{-S[Z',\tilde Z'^\dagger,Z'^\dagger,\tilde Z']}\\
=&
\int [dZd\tilde Z^\dagger dZ^\dagger d\tilde Z]\,e^J\cdot e^{-S[Z,\tilde Z^\dagger,Z^\dagger,\tilde Z]-\delta S},\label{eq:j_ds}
\end{align}
where we use the fact that the path integral is independent of the dummy variables, and $e^J$ is the Jacobian
\begin{align}
e^J=\det\pdv{(Z',\tilde Z'^\dagger)}{(Z,\tilde Z^\dagger)}\det\pdv{(Z'^\dagger,\tilde Z')}{(Z^\dagger,\tilde Z)}.
\end{align}
From here we calculate $J$ and $\delta S$ up to the first order of $\Lambda$. With respect to the variation $\delta Z=i\Lambda Z$,
\begin{align}
\delta S=\int d^6z\,i\Lambda\left[-\frac14\bar D^2\left(Z^\dagger e^VZ\right)+\tilde Z\Phi Z\right].\label{eq:ds}
\end{align}
On the other hand the Jacobian $e^J$ is
\begin{gather}
e^J=\det\left(\pdv{a'}{a}\right),\\
\ket{\Xi}=\sum_na_n\ket{\Xi_n},\quad\ket{\Xi'}=\sum_na'_n\ket{\Xi_n}.
\end{gather}
The coefficient $a_n$ is transformed as
\begin{align}
a'_n&=\braket{\Xi_n}{\Xi'}=\bra{\Xi_n}1+
\begin{pmatrix}
i\Lambda&0\\0&0
\end{pmatrix}
\ket{\Xi}\\
&=\sum_m\bra{\Xi_n}\left(1+
\begin{pmatrix}
i\Lambda&0\\0&0
\end{pmatrix}\right)a_m
\ket{\Xi_m}\\
&\equiv\sum_m\left(\delta_{nm}+C_{nm}\right)a_m.
\end{align}
Thus the Jacobian variation term $J$ is calculated as
\begin{align}
J&=\log\det(1+C)\simeq\Tr C=\sum_nC_{nn}\\
&=\Tr\sum_n
\begin{pmatrix}
\int d^6z\,i\Lambda&0\\0&0
\end{pmatrix}
\ket{\Xi_n}\bra{\Xi_n}\label{eq:before_reg}\\
&\rightarrow\Tr
\begin{pmatrix}
\int d^6z\,i\Lambda&0\\0&0
\end{pmatrix}
\exp\left(\frac P{M^2}\right)
\left.
\begin{pmatrix}
\delta^6(z_L-z'_L)&0\\0&\delta^6(z_R-z'_R)
\end{pmatrix}
\right|_{M\to\infty,z'\to z}\label{eq:after_reg}\\
&=\Tr
\begin{pmatrix}
\int d^6z\,i\Lambda&0\\0&0
\end{pmatrix}
\begin{pmatrix}
\int\frac{d^6w}{4\pi^4}e^{-iwz_L}&0\\0&\int\frac{d^6\bar w}{4\pi^4}e^{-i\bar wz_R}
\end{pmatrix}\nonumber\\
&\qquad\qquad\qquad\qquad\qquad\qquad\times\exp\left(\frac P{M^2}\right)\left.
\begin{pmatrix}
e^{iwz_L}&0\\0&e^{i\bar wz_R}
\end{pmatrix}\right|_{M\to\infty}\\
&=\frac1{8\pi^2}\tr\int d^6z\,i\Lambda X\label{eq:i_lambda_x}
\end{align}
where $\Tr$ denotes full (gauge and matrix) trace, while $\tr$ denotes gauge trace. The divergent sum is regulated by the gauge-invariant and $\mathcal N=2$ supersymmetric regulator $P$ from (\ref{eq:before_reg}) to (\ref{eq:after_reg}), and $X$ is the coefficient of $D^2$ in the chiral-chiral entry\cite{0} of $P^2$, that is,
\begin{align}
X&=-\frac18W^\alpha W_\alpha+\Phi\frac14\bar D^2e^{-V}\Phi^\dagger\frac14e^V+\frac14\bar D^2e^{-V}\Phi^\dagger\frac14e^V\Phi-\frac14\bar D^2e^{-V}\Phi^{\dagger2}\frac14e^V,
\end{align}
where we use
\begin{align}
\frac1{16}\bar D^2e^{-V}D^2e^V&=-\frac12\frac{-1}4\bar D^2e^{-V}D^\alpha e^V\cdot D_\alpha+\cdots\\
&=-\frac12W^\alpha D_\alpha+\cdots,
\end{align}
\begin{align}
\left(-\frac12W^\alpha D_\alpha\right)^2=-\frac14W^\alpha W^\beta D_\alpha D_\beta\to-\frac18W^\alpha W_\alpha.
\end{align}
Thus (\ref{eq:i_lambda_x}) is
\begin{align}
J&\simeq-\frac1{16\pi^2}\tr\int d^6z\,i\Lambda \left[\frac14W^\alpha W_\alpha-\frac14\bar D^2\left(e^{-V}\Phi^\dagger e^V\Phi\right)\right]\nonumber\\
&\qquad+\frac1{32\pi^2}\tr\int d^6z\,i\Lambda\frac{-1}4\bar D^2e^{-V}\Phi^{\dagger2}e^V.
\end{align}
The last term vanishes since
\begin{align}
\tr\int d^6z\,i\Lambda\frac{-1}4\bar D^2e^{-V}\Phi^{\dagger2}e^V&=\tr\int d^8z\,i\Lambda e^{-V}\Phi^{\dagger2}e^V\\
&=\tr\int d^8z\,i\Lambda\Phi^{\dagger2}\\
&=\tr\int d^6\bar z\,i\frac{-1}4D^2\Lambda\Phi^{\dagger2}\\
&=0.
\end{align}
Thus,
\begin{align}
J&\simeq-\frac1{16\pi^2}\tr\int d^6z\,i\Lambda \left[\frac14W^\alpha W_\alpha-\frac14\bar D^2\left(e^{-V}\Phi^\dagger e^V\Phi\right)\right],\label{eq:j_result}
\end{align}
and finally, from (\ref{eq:j_ds}), (\ref{eq:ds}), and (\ref{eq:j_result}),
\begin{align}
\ev{-\frac14\bar D^2\left(Z^\dagger e^VZ\right)+\tilde Z\Phi Z}=-\frac1{16\pi^2}\tr\left[\frac14W^\alpha W_\alpha-\frac14\bar D^2\left(e^{-V}\Phi^\dagger e^V\Phi\right)\right].
\end{align}
Similarly, consider the variation $\delta\tilde Z=-i\Lambda\tilde Z$, $\delta Z^\dagger=-i\Lambda^\dagger Z^\dagger$, and $\delta\tilde Z^\dagger=i\Lambda^\dagger\tilde Z^\dagger$, respectively, to obtain
\begin{align}
\ev{-\frac14\bar D^2\left(\tilde Z e^{-V}\tilde Z^\dagger\right)+\tilde Z\Phi Z}&=-\frac1{16\pi^2}\tr\left[\frac14W^\alpha W_\alpha-\frac14\bar D^2\left(e^{-V}\Phi^\dagger e^V\Phi\right)\right],\\
\ev{-\frac14D^2\left(Z^\dagger e^VZ\right)+Z^\dagger\Phi^\dagger\tilde Z^\dagger}&=-\frac1{16\pi^2}\tr\left[\frac14\bar W_{\dot\alpha}\bar W^{\dot\alpha}-\frac14D^2\left(e^{-V}\Phi^\dagger e^V\Phi\right)\right],\\
\ev{-\frac14D^2\left(\tilde Ze^{-V}\tilde Z^\dagger\right)+Z^\dagger\Phi^\dagger\tilde Z^\dagger}&=-\frac1{16\pi^2}\tr\left[\frac14\bar W_{\dot\alpha}\bar W^{\dot\alpha}-\frac14D^2\left(e^{-V}\Phi^\dagger e^V\Phi\right)\right].
\end{align}
This is the $\mathcal N=2$ chiral anomaly, which we were going to derive.

\section{Summary}\label{sec:summary}
In this paper, we have defined $\mathcal N=2$ supersymmetric and gauge-invariant path integral measure in $\mathcal N=2$ SQCD in terms of $\mathcal N=1$ superfields. As a further consequence, we have derived the $\mathcal N=2$ version of the chiral anomaly in a way that preserves $\mathcal{N}=2$ supersymmetry. The result is
\begin{align}
\ev{-\frac14\bar D^2\left(Z^\dagger e^VZ\right)+\tilde Z\Phi Z}&=-\frac1{16\pi^2}\tr\left[\frac14W^\alpha W_\alpha-\frac14\bar D^2\left(e^{-V}\Phi^\dagger e^V\Phi\right)\right],\\
\ev{-\frac14\bar D^2\left(\tilde Z e^{-V}\tilde Z^\dagger\right)+\tilde Z\Phi Z}&=-\frac1{16\pi^2}\tr\left[\frac14W^\alpha W_\alpha-\frac14\bar D^2\left(e^{-V}\Phi^\dagger e^V\Phi\right)\right],\\
\ev{-\frac14D^2\left(Z^\dagger e^VZ\right)+Z^\dagger\Phi^\dagger\tilde Z^\dagger}&=-\frac1{16\pi^2}\tr\left[\frac14\bar W_{\dot\alpha}\bar W^{\dot\alpha}-\frac14D^2\left(e^{-V}\Phi^\dagger e^V\Phi\right)\right],\\
\ev{-\frac14D^2\left(\tilde Ze^{-V}\tilde Z^\dagger\right)+Z^\dagger\Phi^\dagger\tilde Z^\dagger}&=-\frac1{16\pi^2}\tr\left[\frac14\bar W_{\dot\alpha}\bar W^{\dot\alpha}-\frac14D^2\left(e^{-V}\Phi^\dagger e^V\Phi\right)\right].
\end{align}
This is the $\mathcal N=2$ version of the results by Fujikawa \cite{1} and by Konishi and Shizuya \cite{2,3}.

In addition to this topic, there are also many intriguing topics concerning quantum anomaly or explicit calculation of path integral in $\mathcal N=2$ or more extended supersymmetric theories. Describing those topics by superfields is worth examining. For example, supersymmetric localisation \cite{15,16,17,18}, i.e. a theory considering calculation of path integral around instanton, by superfields is attractive. Another one is checking conformal invariance in quantum $\mathcal N=4$ super-Yang-Mills theory by superfields. Further studies are expected for these topics.

\end{document}